# Field strength-dependent performance variability in deep learning-based analysis of magnetic resonance imaging

Muhammad Ibtsaam Qadir[1], Duane Schonlau[2], Ulrike Dydak[1,2,3], Fiona R. Kolbinger[1,4,#]

[1] Weldon School of Biomedical Engineering, Purdue University, West Lafayette, IN, USA
[2] Department of Radiology and Imaging Sciences, Indiana University School of Medicine, Indianapolis, IN, USA
[3] School of Health Sciences, Purdue University, West Lafayette, IN, USA
[4] Department of Visceral, Thoracic and Vascular Surgery, Medical Faculty and University Hospital Dresden, TUD Dresden University of Technology, Dresden, Germany

# Corresponding author:
Dr. Fiona Kolbinger, Weldon School of Biomedical Engineering, Purdue University, 206 S. Martin Jischke Drive, West Lafayette, IN 47907, USA, email: fkolbing@purdue.edu

## Abstract

**Purpose:** To quantitatively evaluate the impact of MRI scanner magnetic field strength on the performance and generalizability of deep learning-based segmentation algorithms.

**Materials and Methods:** Three publicly available MRI datasets (breast tumor, pancreas, and cervical spine) were stratified by scanner field strength (1.5T vs. 3.0T). For each segmentation task, three nnU-Net-based models were developed: A model trained on 1.5T data only ($model_{1.5T}$), a model trained on 3.0T data only ($model_{3.0T}$), and a model trained on pooled 1.5T and 3.0T data ($model_{combined}$). Each model was evaluated on both 1.5T and 3.0T validation sets. Field-strength-dependent performance differences were further investigated via Uniform Manifold Approximation and Projection (UMAP)-based clustering and radiomic analysis, including 23 first-order and texture features.

**Results:** For breast tumor segmentation, $model_{3.0T}$ (DSC: 0.494 [1.5T] and 0.433 [3.0T]) significantly outperformed $model_{1.5T}$ (DSC: 0.411 [1.5T] and 0.289 [3.0T]) and $model_{combined}$ (DSC: 0.373 [1.5T] and 0.268[3.0T]) on both validation sets ($p<0.0001$). Pancreas segmentation showed similar trends: $model_{3.0T}$ achieved the highest DSC (0.774 [1.5T], 0.840 [3.0T]), while $model_{1.5T}$ underperformed significantly ($p<0.0001$). For cervical spine, models performed optimally on same-field validation sets with minimal cross-field performance degradation (DSC>0.92 for all comparisons). Radiomic analysis revealed moderate field-strength-dependent clustering in soft tissues (silhouette scores 0.23-0.29) but minimal separation in osseous structures (0.12).

**Conclusion:** Magnetic field strength in the training data substantially influences the performance of deep learning-based segmentation models, particularly for soft-tissue structures, important for sensitivity to small lesions. This warrants consideration of magnetic field strength as a confounding factor in studies evaluating AI performance on MRI.

## Introduction

Magnetic resonance (MR) images exhibit statistical distribution shifts arising from variations in scanner manufacturer, acquisition parameters, and imaging protocols (1). These variations introduce domain shifts that can substantially challenge the generalizability of artificial intelligence (AI) models (2–4). Clinical practice routinely involves mixed scanner and acquisition protocols across multiple centers, longitudinal studies, and scanner upgrade cycles, creating a heterogeneous data landscape. As AI models are increasingly adopted for radiology workflow, it is critical to assess their performance across the full spectrum of real-world variations.

Prior studies have shown that both scanner manufacturers (2) and acquisition parameters (5) can affect AI performance. Among the MRI scanner parameters, magnetic field strength is a key determinant of image quality, influencing signal-to-noise ratio (SNR) and tissue contrast via the static magnetic field ($B_0$) (6,7). While higher field strengths have been shown to improve abdominal and neurological imaging (7,8), their impact on AI model performance remains largely unexplored.

This study aims to quantitatively evaluate the impact of MRI magnetic field strength (1.5 T vs. 3.0 T) on the performance and generalizability of deep learning-based segmentation algorithms across three tasks: breast tumor, pancreas, and cervical spine segmentation. By characterizing how field strength influences image characteristics and identifying patterns underlying consequential differences in model performance, this study provides important evidence related to MRI-based AI model training and validation.

## Methods

### Ethics Statement

This retrospective study was conducted in accordance with the Declaration of Helsinki and its subsequent amendments. No identifiable patient data were used in this study; all clinical data used are publicly available. No informed consent was required. The protocol was reviewed and approved by the institutional review board of Purdue University on February 7, 2024 (IRB-2023-1736).

### Datasets

Three publicly available datasets were used to develop and evaluate models across distinct segmentation tasks: breast tumor segmentation (MAMA-MIA) (9), pancreas segmentation (10), and cervical spine segmentation (CSpineSeg) (11). Each dataset was stratified by scanner magnetic field strength (1.5T and 3.0T) to evaluate model performance across different imaging parameters (1.5T vs. 3.0T scanners). Data were obtained from multiple scanner manufacturers and heterogeneous acquisition protocols; other imaging parameters were not standardized. The analysis only focused on magnetic field strength. Exams with incomplete imaging parameter metadata were excluded from analysis. For breast tumor segmentation, first post-contrast DCE-MRI sequences were analyzed. T2-weighted images were used for pancreas segmentation, and sagittal T2-weighted sequences were used for cervical spine segmentation. The MAMA-MIA and pancreas datasets were divided into training and validation cohorts at the center level to prevent data leakage, while CSpineSeg was split using an 80:20 ratio. This stratification resulted in four distinct cohorts per segmentation task, i.e., two training sets (1.5T and 3.0T) and two validation sets (1.5T and 3.0T). The exact distribution of MRIs across datasets and cohorts is provided in Supplementary Methods 1.

## Model Development and Validation

All segmentation models were trained using the nnUNet (12) framework on AMD MI210 GPUs for 250 epochs. A standardized 3D full-resolution nnUNet baseline configuration was used consistently across all experiments to ensure fair comparison. For each segmentation task, three models were developed: one trained exclusively on 1.5T data ($model_{1.5T}$), one trained exclusively on 3.0T data ($model_{3.0T}$), and one trained on the combined 1.5T and 3.0T dataset ($model_{combined}$). This experimental design enabled assessment of whether training on a single field strength versus mixed field strengths affects segmentation performance. Each model was evaluated on both 1.5T and 3.0T validation sets to quantify performance across different scanner field strengths.

## Radiomic Analysis

To assess the impact of field strength on image characteristics, radiomic features were extracted for each segmentation task from both whole MRI volumes and segmented regions of interest (ROIs) for the 1.5T and 3.0T subgroups. A total of 23 features were extracted, comprising 16 first-order intensity-based features and 7 second-order texture features derived from gray-level co-occurrence matrices (GLCM). First-order features included basic statistics, higher-order statistics (skewness, kurtosis), entropy, coefficient of variation, energy, and root mean square intensity. Extracted GLCM features were: contrast, dissimilarity, homogeneity, energy, correlation, and angular second moment (ASM). A full list of features is provided in Supplementary Methods 1. To visualize potential clustering patterns based on field strength, Uniform Manifold Approximation and Projection (UMAP) dimensionality reduction was applied to the extracted feature sets. UMAP was configured with 25 nearest neighbors, a minimum distance of 0.5, and reduction to 2 components to enable visualization of feature space separation between 1.5T and 3.0T acquisitions. The separation was quantified using the silhouette score.

## Statistical Analysis

Segmentation performance was evaluated using the Dice similarity coefficient (DSC) as the primary metric and average symmetric surface distance (ASSD) as the secondary metric. The Wilcoxon signed-rank test was employed to compare model performance on specific validation sets, accounting for paired observations. Performance metrics are reported as mean values with 95% confidence intervals. Radiomic features extracted from 1.5T and 3.0T datasets were compared using the Mann-Whitney U test to assess differences in image characteristics between field strengths. Statistical significance was defined as a *p-value* $< 0.05$. All model development and statistical analyses were performed in Python 3.11.

# Results

## Data Distribution

For breast tumor segmentation, 1,271 DCE-MRI examinations from the MAMA-MIA dataset were analyzed, comprising 851 acquired at 1.5T and 420 at 3.0T. The pancreas segmentation cohort included 218 abdominal MRI examinations (90 at 1.5T and 128 at 3.0T). For cervical spine segmentation of vertebral bodies and intervertebral discs, 1,255 MRI examinations were included, of which 747 were acquired at 1.5T and 508 at 3.0T. The distribution of training and validation cohorts across field strengths for each segmentation task is detailed in Table 1. Patient characteristics have previously been published in (9–11).

**Table 1: Dataset distribution across field strengths and training/validation cohorts.** Numbers indicate the number of MRI examinations for each segmentation task, stratified by scanner field strength (1.5T and 3.0T).

| Magnetic Field Strength | Total Examinations | Training Set (% total) | Validation Set (% total) |
|---|---|---|---|
| **Breast Tumor Segmentation (n = 1,271)** | | | |
| 1.5 T | 851 | 715 (72.96%) | 136 (46.74%) |
| 3.0 T | 420 | 265 (27.04%) | 155 (53.26%) |
| **Pancreas Segmentation (n = 218)** | | | |
| 1.5 T | 90 | 69 (38.98%) | 21 (51.22%) |
| 3.0 T | 128 | 108 (61.02%) | 20 (48.78%) |
| **Cervical Spine Segmentation (n = 1,255)** | | | |
| 1.5 T | 747 | 597 (59.52%) | 150 (59.52%) |
| 3.0 T | 508 | 406 (40.48%) | 102 (40.48%) |

## 3.0T-trained Models Outperform 1.5T-trained Models

For breast tumor segmentation, $model_{3.0T}$ achieved the highest mean DSC across both validation sets: 0.494 (95% CI: 0.450, 0.538) on the 1.5T validation set and 0.433 (95% CI: 0.395, 0.470) on the 3.0T validation set, significantly outperforming both $model_{1.5T}$ (1.5T validation set: $p < 0.0001$, 3.0T validation set: $p < 0.0001$) and $model_{combined}$ (1.5T validation set: $p < 0.0001$, 3.0T validation set: $p < 0.0001$). $Model_{1.5T}$ demonstrated intermediate performance, with mean DSCs of 0.411 (95% CI: 0.369, 0.452) and 0.289 (95% CI: 0.256, 0.322) on the 1.5T and 3.0T validation sets, respectively. $Model_{combined}$ exhibited the lowest performance, achieving DSC values of 0.373 (95% CI: 0.331, 0.414) on the 1.5T validation set and 0.268 (95% CI: 0.236, 0.301) on the 3.0T validation set (Figure 1, Supplementary Table 1). In the volumetric agreement analysis, $model_{3.0T}$ demonstrated the lowest mean bias of all models at +3.694 mL (limits of agreement [LoA]: -86.595 to +93.983 mL) on the 1.5T validation set and +14.026 mL (LoA: -20.371 to +48.422 mL) on the 3.0T validation set, indicating excellent volumetric agreement between segmentation and ground truth for $model_{3.0T}$ (Figure 2A, Supplementary Table 2, Supplementary Figure 1).

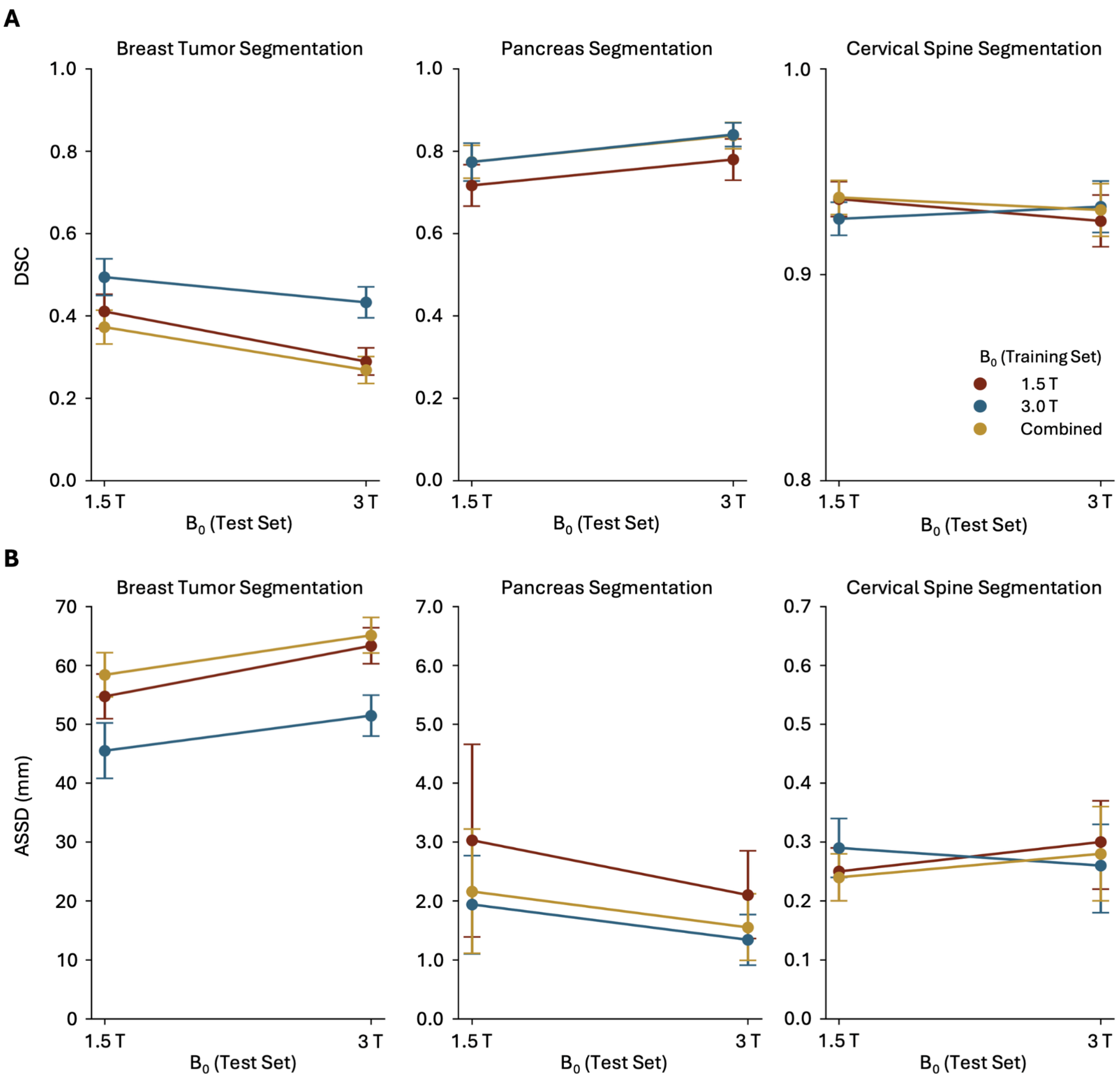

**Figure 1: Model performance across field strengths and training strategies. (A)** Dice similarity coefficient (DSC) and **(B)** average symmetric surface distance (ASSD) across breast tumor, pancreas, and cervical spine segmentation tasks. Three models were evaluated on both 1.5T and 3.0T validation sets: model$_{1.5T}$ (trained on 1.5T data), model$_{3.0T}$ (trained on 3.0T data), and model$_{combined}$ (trained on pooled 1.5T and 3.0T data). Y-axis scales differ across tasks: DSC ranges from 0.8-1.0 for cervical spine and 0-1.0 for other tasks; ASSD ranges from 0-70 mm for breast tumor, 0-7 mm for pancreas, and 0-0.7 mm for cervical spine. Error bars represent 95% confidence intervals. Higher DSC and lower ASSD indicate better performance.

We observed a similar trend for pancreas segmentation. Model$_{3.0T}$ achieved the highest mean DSC on both same-field and cross-field validation sets: 0.774 (95% CI: 0.724, 0.819) on the 1.5T validation set and 0.840 (95% CI: 0.811, 0.869) on the 3.0T validation set. Model$_{combined}$ demonstrated DSC values comparable to those of model$_{3.0T}$, with mean DSCs of 0.774 (95% CI: 0.734, 0.814) and 0.838 (95% CI: 0.806, 0.870) on the 1.5T and 3.0T validation sets, respectively. However, $model_{3.0T}$ exhibited lower ASSD values across both field strengths (Figure 1B). $Model_{1.5T}$ showed significantly lower performance compared to both $model_{3.0T}$ and $model_{combined}$, achieving mean DSC of 0.717 (95% CI: 0.667–0.767) and 0.780 (95% CI: 0.730–0.830) on the 1.5T and 3.0T validation sets, respectively ($p < 0.0001$ for all pairs and validation sets) (Figure 1, Supplementary Table 1, Supplementary Figure 1). The Bland-Altman analysis (Figure 2B, Supplementary Table 2) revealed that $model_{3.0T}$ exhibited a lower mean bias of -22.932 mL (LoA: -77.393 to +31.529 mL) on 1.5T validation data and -6.660 mL (LoA: -65.959 to +52.639 mL) on 3.0T validation data than $model_{1.5T}$ across both fields.

For cervical spine segmentation, models demonstrated excellent performance on same-field validation sets but exhibited modest decreases in cross-field generalization. On the 1.5T validation set, $model_{1.5T}$ and $model_{combined}$ achieved comparable performance with mean DSC values of 0.937 (95% CI: 0.928, 0.945) and 0.938 (95% CI: 0.929, 0.946), respectively, while $model_{3.0T}$ showed slightly lower performance (mean DSC: 0.927, 95% CI: 0.919, 0.935). On the 3.0T validation set, $model_{3.0T}$ achieved the highest mean DSC of 0.933 (95% CI: 0.921, 0.946), followed by $model_{combined}$ (mean DSC: 0.931, 95% CI: 0.919, 0.944) and $model_{1.5T}$ (mean DSC: 0.926, 95% CI: 0.914, 0.939) (Figure 1, Supplementary Table 1). On volumetric agreement analysis, both model1.5T and model3.0T demonstrated excellent volumetric agreement with minimal bias (Supplementary Figure 1). On the 1.5T validation set, $model_{1.5T}$ achieved near-zero bias (+0.042 mL; LoA: -6.833 to +6.916 mL) compared to $model_{3.0T}$ (-1.199 mL; LoA: -8.261 to +5.863 mL). On the 3.0T validation set, $model_{1.5T}$ showed a bias of +0.714 mL (LoA: -4.319 to +5.747 mL) while $model_{3.0T}$ demonstrated -0.484 mL (LoA: -5.131 to +4.164 mL) (Figure 2C, Supplementary Table 2).

These results demonstrate that models trained on 3.0T datasets achieve superior performance compared to those trained on 1.5T. The performance gap is most evident in soft-tissue-dependent applications such as breast tumor and pancreas segmentation. In contrast, segmentation of osseous structures (vertebral bodies and intervertebral discs) showed negligible field-strength-related differences.

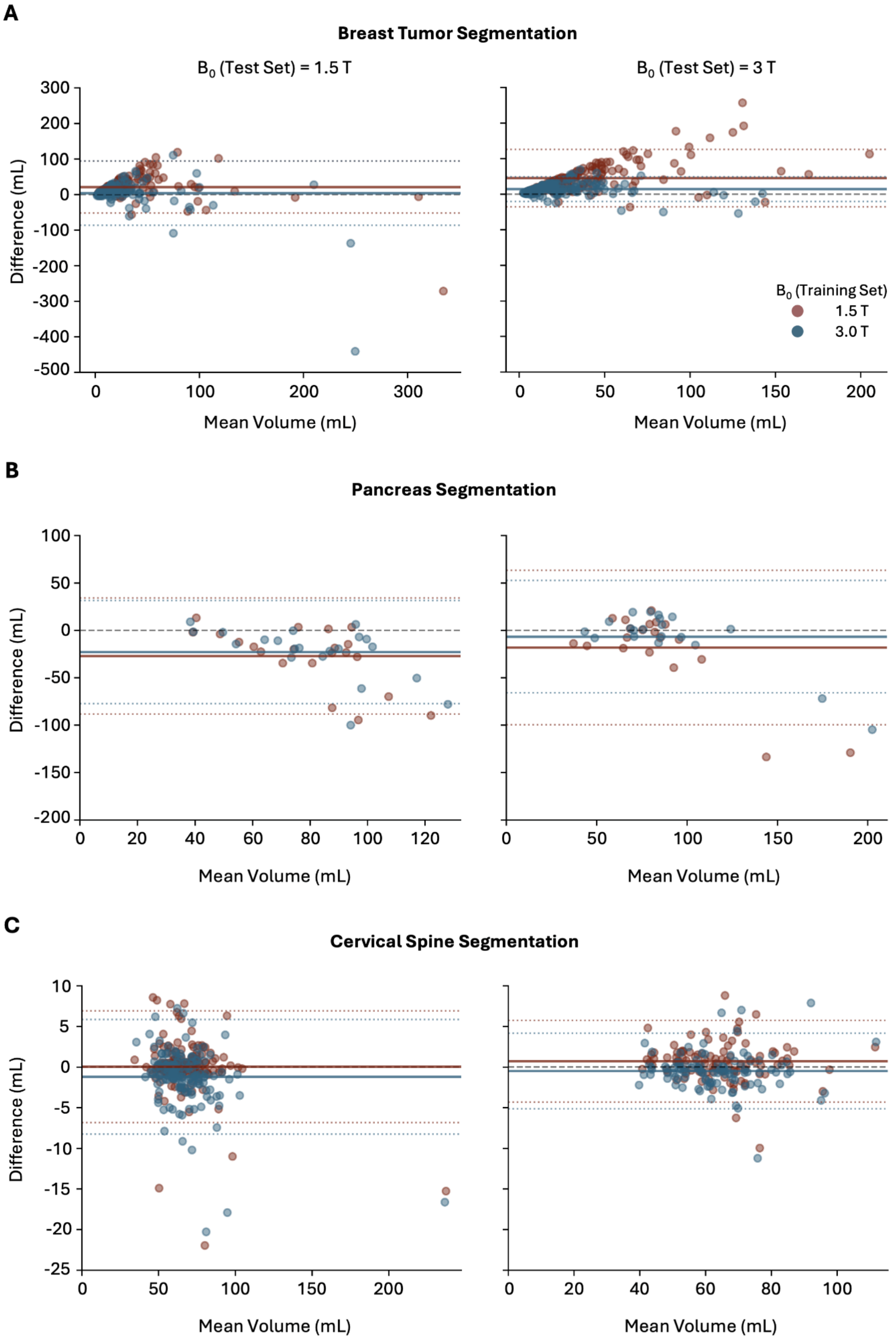

**Figure 2: Volumetric agreement analysis across segmentation tasks and field strengths.** Bland-Altman plots comparing automated segmentation volumes to ground truth for (A) breast tumor, (B) pancreas, and (C) cervical spine. Each task displays results on 1.5T validation sets (left) and 3.0T validation sets (right). The x-axis represents the mean volume of ground truth and predicted segmentation, while the y-axis shows the volume difference. Solid

horizontal lines indicate mean bias, dashed lines indicate 95% limits of agreement (±1.96 standard deviation), and colors distinguish models trained on different field strengths ($model_{1.5T}$ and $model_{3.0T}$). Narrower limits of agreement indicate more consistent volumetric measurements. X- and y-axis scales differ across tasks.

## Association between Radiomic Features and Magnetic Field Strength of the Scanner

Radiomic feature extraction and UMAP dimensionality reduction showed varying degrees of field strength-dependent clustering across segmentation tasks, quantified using silhouette scores. ROI-based analysis revealed moderate clustering by field strength for breast tumor (s = 0.227) and pancreas (s = 0.287), while vertebral bodies demonstrated weaker separation (s = 0. 117) (Figure 3A). Whole-volume analysis showed variable field strength discrimination across anatomical regions. DCE-breast MRI exhibited minimal separation (s = 0.077), abdominal MRI showed moderate clustering (s = 0.259), and cervical spine MRI demonstrated strong field strength-dependent clustering (s = 0.305) (Figure 3B).

When comparing the individual quantitative image characteristics between 1.5T and 3.0T data across all use cases, mean intensity and intensity standard deviation differed consistently between field strengths for both whole-volume and ROI-based analyses. 3.0T MRIs demonstrated markedly higher signal intensities and greater intensity dispersion in soft-tissue structures (Figure 4A and 4B, Supplementary Table 3). This reflects the expected gain in SNR and dynamic range at higher field strength. ROI-specific texture analysis revealed that GLCM homogeneity was significantly increased at 3.0T in the breast tumors, whereas homogeneity differences were negligible in the other anatomical regions (Figure 4C, Supplementary Table 3). In contrast, GLCM dissimilarity differed significantly across all ROIs, with higher dissimilarity values observed in the 1.5T subgroup (Figure 4D, Supplementary Table 3).

These findings reveal systematic differences in image characteristics between 1.5T and 3.0T acquisitions, with the magnitude of separation varying across anatomical regions and potentially contributing to the observed field strength-specific model performance.

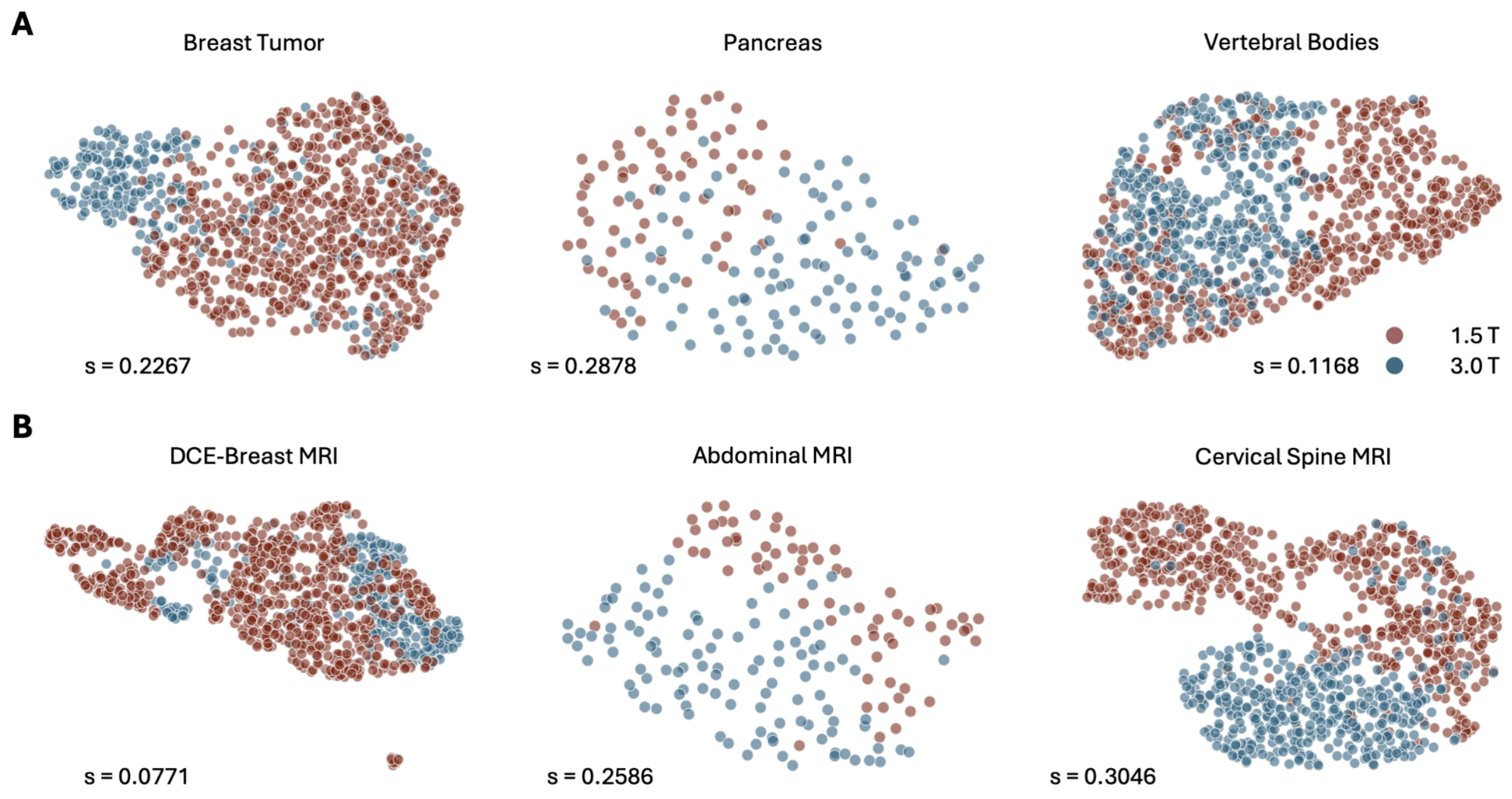

**Figure 3: UMAP visualization of radiomic feature clustering by scanner field strength.** Two-dimensional UMAP projections of extracted radiomic features (first-order intensity statistics and GLCM texture features) for **(A)** segmented regions of interest and **(B)** whole MRI volumes across breast tumor, pancreas, and cervical spine datasets. Separation of points by field strength indicates systematic differences in quantitative image-derived tissue characteristics between 1.5T and 3.0T acquisitions, even for the same anatomic regions. Silhouette scores quantify the degree of separation between field strengths, with higher values indicating stronger clustering.

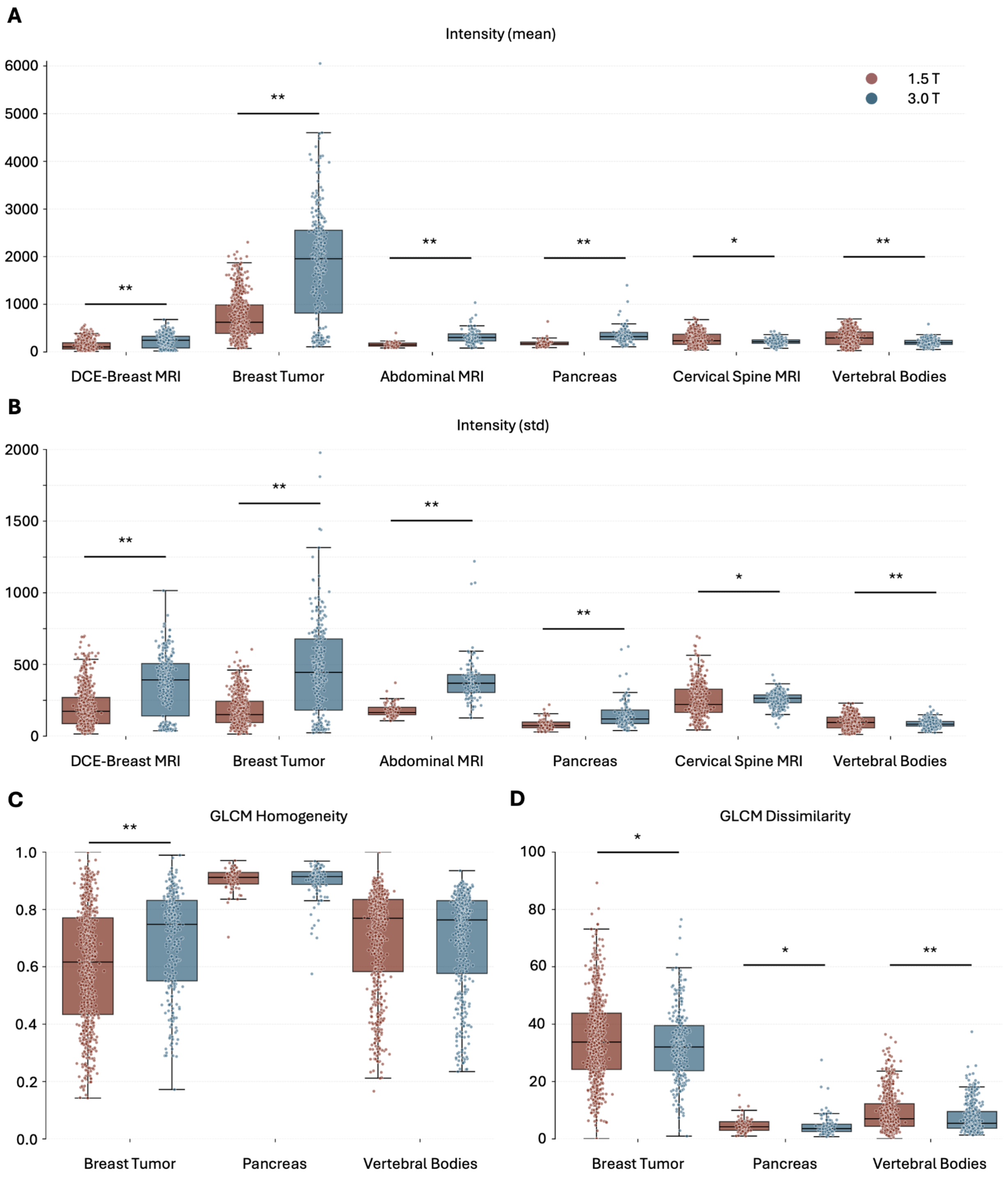

**Figure 4: Field strength-dependent differences in individual quantitative image characteristics.** Whole-volume and ROI-based analyses demonstrate consistently higher **(A)** mean intensity and **(B)** intensity standard deviation in 3.0T examinations compared to 1.5T across soft-tissue dominant use cases. **(C)** GLCM homogeneity is significantly increased at 3.0T in the breast tumor cohort, with minimal differences observed in other anatomical regions. **(D)** GLCM dissimilarity is significantly higher in the 1.5T subgroup across all ROIs. Supplementary Table 3 provides full statistical comparisons. ** indicates $p < 0.0001$; * indicates $p < 0.05$.

## Discussion

Understanding the generalizability of AI models across varying image acquisition parameters is critical for effective deployment in clinical settings. Recent works have established that scanner manufacturers, acquisition parameters, and protocols affect deep learning model performance and generalizability (2,5). This study focused specifically on the influence of magnetic field strength ($B_0$), a primary determinant of SNR in MRI (6,7).

Our results reveal that models trained exclusively on 3.0T data significantly outperformed those trained on 1.5T data across soft-tissue-dependent applications, such as breast tumor and pancreas segmentation ($p < 0.0001$). In contrast, we observed minimal field strength-dependent differences for segmentation of osseous structures such as vertebral bodies and intervertebral discs. These effects extend beyond segmentation metrics to fundamental quantitative imaging characteristics, as demonstrated through UMAP and radiomic feature analysis. The observed clustering patterns (silhouette scores ranging from 0.077 to 0.305) confirm the presence of systematic differences in image properties across field strengths, with the magnitude varying by anatomical region and tissue composition. This trend can be attributed to the inherently higher SNR at increased field strength, as it is approximately linearly related with $B_0$ (13), resulting in images with greater signal intensity and enhanced tissue differentiation (6,7). Our radiomic analysis corroborates this, demonstrating that 3.0T acquisitions exhibit significantly higher mean intensities compared to 1.5T, particularly in soft tissues (e.g., abdomen) where the SNR advantage is most impactful (7). The increased dynamic range at 3.0T provides deep learning models with richer input information, enabling more precise feature extraction and boundary delineation. Moreover, our combined training approach did not consistently outperform field-specific models, rather underperformed relative to both 1.5T and 3.0T-specific models for breast tumor segmentation. This indicates that simply pooling data across field strengths may introduce feature noise that hinders model optimization. Strategies such as domain adaptation, multi-task learning, or explicit conditioning on field strength may be required to effectively mitigate this domain shift (14).

Our findings have important implications for clinical AI deployment and multicenter research studies. First, multi-center clinical trials involving sites with mixed data from 1.5T and 3.0T scanners must account for field strength as a potential source of performance variability. Prior studies have not consistently reported scanner magnetic field strength (15) or conducted sub-stratified analyses (16,17). The performance differences observed in the present work suggest that such stratified analyses are warranted. Moreover, ongoing and future clinical trials evaluating AI applications for MRI should consider this as a potential confounder in study design and analysis (18). Second, in scenarios where MRI segmentation models are deployed clinically, scanner transitions from 1.5T to 3.0T systems necessitate careful model revalidation, as models trained on legacy 1.5T data may underperform on newer 3.0T exams. Third, clinical deployment strategies should incorporate field-strength-aware validation protocols, with performance benchmarks established separately for each field strength. Finally, AI model documentation should include the training set's field strength as essential metadata, alongside other acquisition parameters.

This study has several limitations. The analysis was limited to three anatomical regions and used unpaired 1.5T and 3.0T cohorts. Therefore, extension to other tissues is needed. We did not evaluate the potential effects of additional parameters (e.g., TE, TR, flip angle), which may interact with field-strength effects. Finally, our combined training strategy did not incorporate domain adaptation approaches that could effectively harmonize multi-field datasets. Future work should systematically evaluate these factors across a wider array of imaging tasks and protocols and determine minimal data requirements for field-specific fine-tuning in real-world clinical settings.

In conclusion, magnetic field strength is a key determinant of AI model performance, with the most pronounced effects observed in soft-tissue tasks. These findings highlight the need for field-strength-aware training and validation in clinical AI development. As AI becomes increasingly integrated into radiological workflows, consideration of acquisition parameters will be essential for ensuring generalizable model deployment in clinical settings.

## Acknowledgments and Funding

FRK receives support from the German Cancer Research Center (CoBot 2.0), the Joachim Herz Foundation (Add-On Fellowship for Interdisciplinary Life Science), the Central Indiana Corporate Partnership AnalytiXIN Initiative, the Evan and Sue Ann Werling Pancreatic Cancer Research Fund, and the Indiana Clinical and Translational Sciences Institute (EPAR4157) funded, in part, by Grant Number UM1TR004402 from the National Institutes of Health, National Center for Advancing Translational Sciences, Clinical and Translational Sciences Award. The content is solely the responsibility of the authors and does not necessarily represent the official views of the National Institutes of Health.

## Author Contributions

Conceptualization: MIQ, DS, FRK; Methodology: MIQ, FRK; Software: MIQ; Validation: MIQ, FRK; Formal analysis: MIQ; Investigation: MIQ, DS, FRK; Resources: FRK; Data Curation: MIQ; Writing - Original Draft: MIQ, FRK; Writing - Review & Editing: MIQ, DS, UD, FRK; Visualization: MIQ, FRK; Supervision: FRK; Project administration: FRK; Funding acquisition: FRK.

## Competing Interests

FRK declares advisory roles for Radical Healthcare, USA; and the Surgical Data Science Collective, USA. All other authors declare no competing interests.

# Field strength-dependent performance variability in deep learning-based analysis of magnetic resonance imaging

## Supplementary Materials

# Supplementary Methods 1

## Datasets

Three public datasets were used for breast tumor (MAMA-MIA) (9), pancreas (10), and cervical spine (CSpineSeg) (11) segmentation. Each was split by scanner field strength (1.5T vs. 3.0T), yielding four cohorts per task: 1.5T training, 3.0T training, 1.5T validation, and 3.0T validation. The exact distribution of MRIs across datasets and cohorts is provided below.

Breast Tumor Segmentation (MAMA-MIA). To reduce dataset heterogeneity, only axial DCE-MRI series were included, resulting in the exclusion of 235 studies (final n = 1,271). Exams from two centers, ISPY2 (n = 980) and DUKE (n = 291), were used for the model development. ISPY2 served as the training cohort, while DUKE was reserved exclusively for external validation.

Pancreas Segmentation. Images lacking scanner field-strength metadata were excluded, resulting in a final cohort of 218 examinations. Data from four centers (NYU, NWU, MCA, and MCF) were included. Examinations from NYU and NWU constituted the training cohort (n = 177), whereas those from MCA and MCF comprised the validation cohort (n = 41).

Cervical Spine Segmentation. All 1,255 examinations in the dataset were included for model development. Because the dataset originated from a single center, an 80:20 split was applied, yielding 1,003 examinations for training and 252 for validation.

## Radiomic Analysis

First-order intensity and texture features were extracted from the ROIs and whole MRI volumes. For each, voxel intensities within the ROI and for whole MRI volume were isolated, and first-order statistics were computed, including mean, median, standard deviation, minimum, maximum, range, 10th/25th/75th/90th percentiles, skewness, kurtosis, entropy, coefficient of variation, energy, and root-mean-square intensity. These features quantified the distribution and variability of signal intensities within the whole volume and the ROI. Texture features were derived using gray-level co-occurrence matrix (GLCM) analysis. Prior to GLCM computation, voxel intensities were clipped to the 1st–99th percentile range and linearly scaled to 8-bit (0–255) to reduce noise. A bounding box surrounding the ROI was determined, and the mid-axial slice of this volume was used for texture analysis for computational efficiency. GLCMs were generated using multiple distances (1, 2, and 3 voxels) and four angular directions (0°, 45°, 90°, and 135°). For each GLCM, contrast, dissimilarity, homogeneity, energy, correlation, and angular second moment (ASM) were calculated. The mean value across all distances and directions was recorded for each texture descriptor.

## Supplementary Table 1

**Supplementary Table 1: Performance metrics of across field strengths and training strategies.** For each evaluation metric, the mean value and corresponding 95% confidence interval are reported. For cervical spine segmentation, performance metrics were averaged across both the vertebral bodies and the intervertebral discs to provide a comprehensive assessment of the model's segmentation performance. P-values were calculated using the Wilcoxon signed-rank test, with the Dice score designated as the primary metric for statistical comparison.

| Validation Set Field | DSC | | | ASSD (mm) | | | p-value |
|---|---|---|---|---|---|---|---|
| | **$Model_{1.5T}$** | **$Model_{3.0T}$** | **$Model_{combined}$** | **$Model_{1.5T}$** | **$Model_{3.0T}$** | **$Model_{combined}$** | |
| **Breast Tumor Segmentation** | | | | | | | |
| 1.5 T | 0.4105 [0.3690, 0.452] | 0.4939 [0.4495, 0.5383] | 0.3725 [0.3313, 0.4136] | 54.72 [50.92, 58.53] | 45.50 [40.79, 50.22] | 58.40 [54.64, 62.15] | $Model_{1.5T}$ vs $Model_{3.0T}$: <0.0001<br>$Model_{1.5T}$ vs $Model_{combined}$: <0.0001<br>$Model_{3.0T}$ vs $Model_{combined}$: <0.0001 |
| 3.0 T | 0.2889 [0.2559, 0.3220] | 0.4326 [0.3948, 0.4704] | 0.2684 [0.2356, 0.3012] | 63.33 [60.27, 66.39] | 51.46 [47.98, 54.94] | 65.12 [62.09, 68.16] | $Model_{1.5T}$ vs $Model_{3.0T}$: <0.0001<br>$Model_{1.5T}$ vs $Model_{combined}$: <0.0001<br>$Model_{3.0T}$ vs $Model_{combined}$: <0.0001 |
| **Pancreas Segmentation** | | | | | | | |
| 1.5 T | 0.7169 [0.6666, 0.7671] | 0.7737 [0.7278, 0.8195] | 0.7741 [0.7343, 0.8140] | 3.03 [1.39, 4.66] | 1.94 [1.10, 2.77] | 2.16 [1.11, 3.22] | $Model_{1.5T}$ vs $Model_{3.0T}$: <0.0001<br>$Model_{1.5T}$ vs $Model_{combined}$: <0.0001<br>$Model_{3.0T}$ vs $Model_{combined}$: 0.9457 |
| 3.0 T | 0.7797 [0.7296, 0.8298] | 0.8403 [0.8114, 0.8691] | 0.8379 [0.8057, 0.8701] | 2.10 [1.36, 2.85] | 1.34 [0.91, 1.77] | 1.55 [0.99, 2.12] | $Model_{1.5T}$ vs $Model_{3.0T}$: <0.0001<br>$Model_{1.5T}$ vs $Model_{combined}$: <0.0001<br>$Model_{3.0T}$ vs $Model_{combined}$: 0.7285 |
| **Cervical Spine Segmentation** | | | | | | | |
| 1.5 T | 0.9367 [0.9282, 0.9452] | 0.9271 [0.9191, 0.9351] | 0.9375 [0.9292, 0.9459] | 0.25 [0.20, 0.29] | 0.29 [0.24, 0.34] | 0.24 [0.20, 0.28] | $Model_{1.5T}$ vs $Model_{3.0T}$: <0.0001<br>$Model_{1.5T}$ vs $Model_{combined}$: 0.0377<br>$Model_{3.0T}$ vs $Model_{combined}$: <0.0001 |
| 3.0 T | 0.9066 [0.8909, 0.9223] | 0.9330 [0.9205, 0.9455] | 0.9314 [0.9186, 0.9442] | 0.30 [0.22, 0.37] | 0.26 [0.18, 0.33] | 0.28 [0.20, 0.36] | $Model_{1.5T}$ vs $Model_{3.0T}$: <0.0001<br>$Model_{1.5T}$ vs $Model_{combined}$: <0.0001<br>$Model_{3.0T}$ vs $Model_{combined}$: <0.0941 |

## Supplementary Figure 1

**Supplementary Figure 1: Qualitative comparison of segmentation performance across scanner field strengths and training strategies.** Ground-truth annotation masks and corresponding predictions from $model_{1.5T}$ and $model_{3.0T}$ data are shown for **(A)** breast tumor, **(B)** pancreas, and **(C)** cervical spine (vertebral bodies) segmentation.

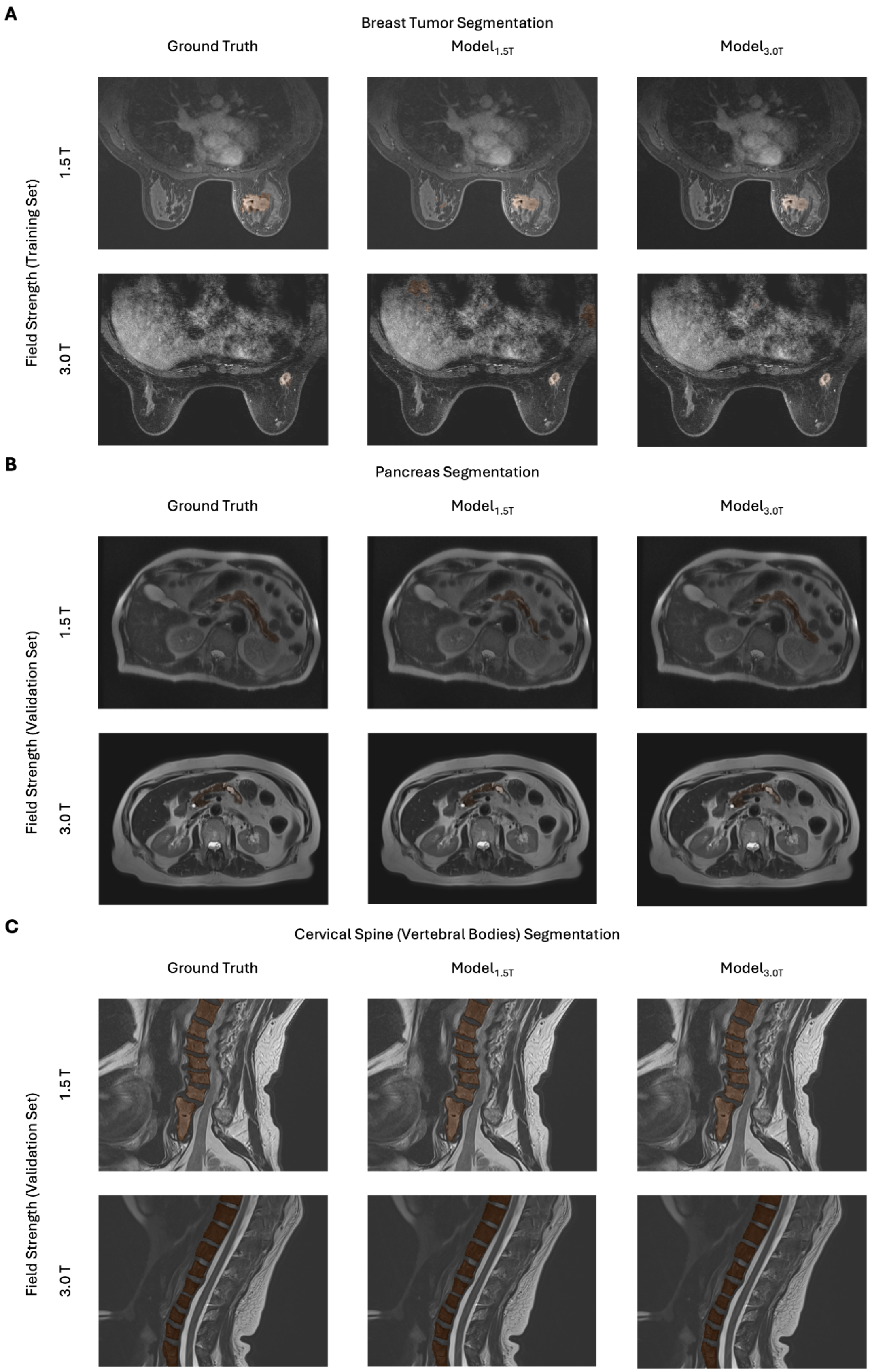

## Supplementary Table 2

**Supplementary Table 2: Volumetric agreement analysis.** For each comparison, the mean bias and corresponding limits of agreement are reported based on Bland–Altman analysis. All volumetric measurements are expressed in milliliters (mL).

| Validation Set Field | Model1.5T | Model3.0T |
|---|---|---|
| **Breast Tumor Segmentation** | | |
| 1.5 T | +20.982 [-52.211, +94.175] | +3.694 [-86.595, +93.983] |
| 3.0 T | +45.023 [-35.308, +125.354] | +14.026 [-20.371, +48.422] |
| **Pancreas Segmentation** | | |
| 1.5 T | -27.139 [-88.387, +34.110] | -22.932 [-77.393, +31.529] |
| 3.0 T | -18.159 [-99.663, +63.344] | -6.660 [-65.959, +52.639 |
| **Cervical Spine Segmentation** | | |
| 1.5 T | +0.042 [-6.833, +6.916] | -1.199 [-8.261, +5.863] |
| 3.0 T | +0.714 [-4.319, +5.747] | -0.484 [-5.131, +4.164] |

## Supplementary Table 3

**Supplementary Table 3: Radiomic feature analysis.** A total of 23 first-order intensity and GLCM-based radiomic features were extracted from both the whole MRI volume and the defined ROI. For breast tumor segmentation, the ROI corresponded to the breast tumor; for pancreas segmentation, the ROI was the pancreas; and for cervical spine segmentation, the ROI comprised the vertebral bodies. Results are reported as mean ± standard deviation. P-values were calculated using the Mann–Whitney U test.

| Feature | Whole MRI Volume | | p-value | Region of Interest | | p-value |
|---|---|---|---|---|---|---|
| | 1.5 T | 3.0 T | | 1.5 T | 3.0 T | |
| **Breast Tumor Segmentation** | | | | | | |
| Intensity mean | 136.7856 ± 96.4847 | 233.1692 ± 141.6296 | < 0.0001 | 729.0425 ± 431.6899 | 1820.1134 ± 1145.3813 | < 0.0001 |
| Intensity median | 47.0657 ± 53.9671 | 68.1610 ± 61.1574 | < 0.0001 | 735.1629 ± 442.2827 | 1830.0084 ± 1161.1860 | < 0.0001 |
| Intensity SD | 196.8279 ± 136.7942 | 357.3350 ± 209.9171 | < 0.0001 | 177.3509 ± 108.2240 | 459.9249 ± 327.4753 | < 0.0001 |
| Intensity minimum | -0.0266 ± 0.3230 | 0.0604 ± 0.4221 | 0.0001 | 113.7790 ± 130.8406 | 254.1964 ± 305.8872 | < 0.0001 |
| Intensity maximum | 1974.4435 ± 1130.7931 | 4628.1763 ± 3464.2001 | < 0.0001 | 1332.7083 ± 767.3538 | 3396.9262 ± 2284.3576 | < 0.0001 |
| Intensity range | 1974.4701 ± 1130.8055 | 4628.1159 ± 3464.1579 | < 0.0001 | 1218.9293 ± 726.8157 | 3142.7298 ± 2188.3533 | < 0.0001 |
| Intensity 10th percentile | 1.6294 ± 3.3952 | 0.7208 ± 1.5803 | 0.0039 | 496.7798 ± 308.6611 | 1222.9180 ± 782.6386 | < 0.0001 |
| Intensity 25th percentile | 4.7507 ± 6.6894 | 5.2953 ± 6.5490 | 0.7036 | 606.2542 ± 369.3818 | 1492.2543 ± 952.2500 | < 0.0001 |
| Intensity 75th percentile | 201.2790 ± 150.2650 | 329.4751 ± 210.7523 | < 0.0001 | 855.1917 ± 504.9830 | 2147.1398 ± 1356.7371 | < 0.0001 |
| Intensity 90th percentile | 388.3089 ± 285.1593 | 691.4020 ± 424.1728 | < 0.0001 | 952.0589 ± 552.3341 | 2406.3350 ± 1528.7369 | < 0.0001 |
| Skewness | 2.3678 ± 0.7243 | 2.5417 ± 0.8566 | 0.0270 | -0.1076 ± 0.4371 | -0.0863 ± 0.4503 | 0.6064 |
| Kurtosis | 8.0355 ± 5.9785 | 10.1354 ± 10.5865 | 0.0533 | 0.1110 ± 0.8146 | -0.0378 ± 0.6611 | 0.0141 |
| Entropy | 1.9522 ± 0.3300 | 1.8429 ± 0.3529 | 0.0001 | 3.3448 ± 0.1894 | 3.3495 ± 0.1682 | 0.8560 |
| Coefficient of Variance | 1.4647 ± 0.2348 | 1.5350 ± 0.2271 | < 0.0001 | 0.2500 ± 0.0702 | 0.2531 ± 0.0730 | 0.7377 |
| Intensity energy | 560496834226.4570 ± 812054467376.9022 | 2127677280264.4878 ± 2025289224029.5547 | < 0.0001 | 19873052865.7578 ± 49992277508.5401 | 184748828740.7251 ± 462826544241.9259 | < 0.0001 |
| Intensity root mean square | 240.2911 ± 166.5331 | 427.4275 ± 251.9596 | < 0.0001 | 751.6904 ± 442.7003 | 1881.5947 ± 1184.4926 | < 0.0001 |
| GLCM mean contrast | 629.7457 ± 402.2565 | 667.2537 ± 341.6079 | 0.0083 | 1881.5947 ± 1184.4926 | 6336.3095 ± 3014.5345 | 0.0427 |
| GLCM contrast SD | 240.1918 ± 140.5146 | 292.6527 ± 129.1451 | < 0.0001 | 2056.1488 ± 1057.9464 | 2236.3290 ± 1000.6955 | 0.0092 |
| GCLM dissimilarity | 13.4138 ± 4.9124 | 13.7819 ± 3.7604 | 0.1251 | 34.6847 ± 14.5147 | 32.4410 ± 12.4031 | 0.0380 |
| GLCM homogeneity | 0.2554 ± 0.0838 | 0.2548 ± 0.0613 | 0.9925 | 0.6026 ± 0.2032 | 0.6895 ± 0.1767 | < 0.0001 |
| GLCM Energy | 0.1303 ± 0.0805 | 0.1293 ± 0.0600 | 0.4028 | 0.5220 ± 0.1868 | 0.5636 ± 0.1631 | 0.0005 |
| GLCM correlation | 0.8946 ± 0.0536 | 0.8946 ± 0.0536 | 0.1233 | 0.6825 ± 0.1115 | 0.7228 ± 0.1003 | < 0.0001 |
| GLCM ASM | 0.0235 ± 0.0399 | 0.0203 ± 0.0181 | 0.4062 | 0.3091 ± 0.2050 | 0.3453 ± 0.1875 | 0.0005 |
| **Pancreas Segmentation** | | | | | | |
| Intensity mean | 156.5430 ± 48.4317 | 322.9109 ± 143.1146 | < 0.0001 | 187.8596 ± 73.3355 | 355.4655 ± 180.2197 | < 0.0001 |
| Intensity median | 85.6087 ± 58.2116 | 144.6852 ± 119.4298 | 0.0008 | 168.2609 ± 66.9189 | 317.1852 ± 154.1947 | < 0.0001 |
| Intensity SD | 178.8652 ± 47.1798 | 390.0407 ± 159.7711 | < 0.0001 | 81.3881 ± 37.2743 | 150.8485 ± 101.4084 | < 0.0001 |
| Intensity minimum | 0.0000 ± 0.0000 | 0.0000 ± 0.0000 | 1.0000 | 17.3478 ± 25.5008 | 28.7407 ± 33.4740 | 0.0601 |
| Intensity maximum | 1288.4348 ± 622.9302 | 2361.2500 ± 1160.1829 | < 0.0001 | 753.6087 ± 218.5904 | 1457.7407 ± 639.6544 | < 0.0001 |
| Intensity range | 1288.4348 ± 622.9302 | 2361.2500 ± 1160.1829 | < 0.0001 | 736.2609 ± 213.6229 | 1429.0000 ± 639.7737 | < 0.0001 |
| Intensity 10th percentile | 2.5362 ± 4.2688 | 1.5556 ± 2.6872 | 0.9277 | 115.5348 ± 48.3901 | 218.5833 ± 106.9270 | < 0.0001 |
| Intensity 25th percentile | 10.3478 ± 14.8560 | 10.8889 ± 24.1956 | 0.6439 | 138.0290 ± 55.8730 | 258.1505 ± 124.4166 | < 0.0001 |
| Intensity 75th percentile | 266.0145 ± 95.3396 | 559.5556 ± 288.4920 | < 0.0001 | 215.9275 ± 90.0360 | 415.1458 ± 235.4679 | < 0.0001 |
| Intensity 90th percentile | 427.0290 ± 109.4916 | 944.0556 ± 387.3151 | < 0.0001 | 282.6594 ± 126.5862 | 536.4343 ± 301.0351 | < 0.0001 |
| Skewness | 1.1990 ± 0.5666 | 1.2040 ± 0.5046 | 0.9796 | 2.1892 ± 0.9260 | 2.2343 ± 0.8731 | 0.7328 |
| Kurtosis | 1.0986 ± 2.1243 | 0.7124 ± 1.7880 | 0.5149 | 9.5488 ± 8.4205 | 10.5644 ± 8.9326 | 0.4621 |
| Entropy | 2.5296 ± 0.3126 | 2.5246 ± 0.3076 | 0.7328 | 2.7784 ± 0.3310 | 2.7247 ± 0.3068 | 0.2575 |
| Coefficient of Variance | 1.1727 ± 0.1943 | 1.2540 ± 0.2117 | 0.0234 | 0.4356 ± 0.1309 | 0.4077 ± 0.1085 | 0.1126 |
| Intensity energy | 215093435630.9420 ± 165446811348.9115 | 1827554888167.7500 ± 4932774835313.7227 | < 0.0001 | 503825377.4638 ± 764369674.9415 | 3383803103.6759 ± 10193545652.2420 | < 0.0001 |
| Intensity root mean square | 238.5411 ± 64.5154 | 508.0861 ± 210.3426 | < 0.0001 | 205.9316 ± 79.1686 | 387.9716 ± 203.3191 | < 0.0001 |
| GLCM mean contrast | 1292.3174 ± 501.3492 | 1145.1624 ± 329.9650 | 0.0367 | 392.0734 ± 308.0642 | 313.7455 ± 390.4716 | 0.0057 |
| GLCM contrast SD | 442.8388 ± 159.5853 | 404.0902 ± 107.8390 | 0.1047 | 87.8584 ± 82.8439 | 77.0702 ± 125.3676 | 0.0203 |
| GCLM dissimilarity | 20.3978 ± 5.0620 | 18.3537 ± 3.5715 | 0.0030 | 4.7334 ± 2.5082 | 4.4259 ± 3.6142 | 0.0302 |
| GLCM homogeneity | 0.2092 ± 0.0701 | 0.2789 ± 0.0729 | < 0.0001 | 0.9054 ± 0.0394 | 0.8985 ± 0.0607 | 0.7260 |
| GLCM Energy | 0.0958 ± 0.0633 | 0.1583 ± 0.0731 | < 0.0001 | 0.9022 ± 0.0404 | 0.8944 ± 0.0632 | 0.8017 |
| GLCM correlation | 0.8450 ± 0.0414 | 0.8493 ± 0.0428 | 0.3320 | 0.5496 ± 0.0873 | 0.5237 ± 0.0979 | 0.0948 |
| GLCM ASM | 0.0131 ± 0.0177 | 0.0304 ± 0.0255 | < 0.0001 | 0.8157 ± 0.0695 | 0.8042 ± 0.1034 | 0.8040 |
| **Cervical Spine Segmentation** | | | | | | |
| Intensity mean | 265.9612 ± 133.4358 | 217.3525 ± 56.0574 | 0.0001 | 295.8379 ± 156.9872 | 201.1666 ± 62.7821 | < 0.0001 |
| Intensity median | 183.3652 ± 108.5204 | 117.0616 ± 48.7354 | < 0.0001 | 299.5519 ± 161.5658 | 200.0653 ± 65.2055 | < 0.0001 |
| Intensity SD | 246.4166 ± 107.4608 | 257.0102 ± 47.0975 | 0.0001 | 96.8877 ± 45.6741 | 86.0413 ± 25.8765 | 0.0020 |
| Intensity minimum | 0.0000 ± 0.0000 | 0.0000 ± 0.0000 | 1.0000 | 33.3283 ± 33.1819 | 0.0000 ± 0.0000 | < 0.0001 |
| Intensity maximum | 2036.5578 ± 1234.9815 | 2020.7586 ± 445.5192 | 0.0001 | 850.4858 ± 400.1929 | 794.8079 ± 217.8453 | 0.1568 |
| Intensity range | 2036.5578 ± 1234.9815 | 2020.7586 ± 445.5192 | 0.0001 | 817.1575 ± 374.8763 | 794.8079 ± 217.8453 | 0.9072 |
| Intensity 10th percentile | 44.0570 ± 34.7164 | 8.9384 ± 4.8872 | < 0.0001 | 164.5797 ± 96.5070 | 92.4961 ± 32.7556 | < 0.0001 |
| Intensity 25th percentile | 68.2027 ± 48.8181 | 29.0690 ± 18.3955 | < 0.0001 | 235.2785 ± 130.7953 | 147.6121 ± 48.5382 | < 0.0001 |
| Intensity 75th percentile | 389.9296 ± 199.7415 | 307.9612 ± 95.3783 | < 0.0001 | 354.8291 ± 186.1915 | 248.2026 ± 78.4333 | < 0.0001 |
| Intensity 90th percentile | 633.7404 ± 287.6389 | 612.9466 ± 153.2706 | 0.4982 | 411.6424 ± 208.6303 | 303.5741 ± 92.8744 | < 0.0001 |
| Skewness | 1.3691 ± 0.4436 | 1.6920 ± 0.4590 | < 0.0001 | 0.2364 ± 0.4672 | 0.5875 ± 0.5640 | < 0.0001 |
| Kurtosis | 1.7593 ± 2.2328 | 2.7945 ± 2.4828 | < 0.0001 | 0.9488 ± 1.3655 | 2.1182 ± 2.5619 | < 0.0001 |
| Entropy | 2.7680 ± 0.2923 | 2.5531 ± 0.2830 | < 0.0001 | 3.1385 ± 0.1777 | 3.0371 ± 0.1967 | < 0.0001 |
| Coefficient of Variance | 0.9784 ± 0.1599 | 1.2116 ± 0.1590 | < 0.0001 | 0.3465 ± 0.0644 | 0.4324 ± 0.0555 | < 0.0001 |
| Intensity energy | 643351657937.2614 ± 569235387774.0526 | 198769025791.3350 ± 99154254232.0401 | < 0.0001 | 8979345676.9648 ± 8343930727.3119 | 1616370155.5616 ± 1129030293.3641 | < 0.0001 |
| Intensity root mean square | 363.5147 ± 169.3083 | 337.2386 ± 70.1855 | 0.8943 | 311.6436 ± 162.8383 | 219.0277 ± 67.1486 | < 0.0001 |
| GLCM mean contrast | 2368.9194 ± 905.0074 | 2815.8691 ± 870.1927 | < 0.0001 | 556.3241 ± 535.0386 | 309.2187 ± 312.6600 | < 0.0001 |
| GLCM contrast SD | 487.0466 ± 206.9114 | 663.3521 ± 230.6846 | < 0.0001 | 150.1396 ± 153.6513 | 82.3682 ± 87.3999 | < 0.0001 |
| GCLM dissimilarity | 31.6217 ± 7.1078 | 33.3588 ± 5.6569 | 0.0012 | 9.2912 ± 6.7440 | 7.2226 ± 5.0439 | < 0.0001 |
| GLCM homogeneity | 0.0756 ± 0.0236 | 0.0780 ± 0.0152 | < 0.0001 | 0.6964 ± 0.1811 | 0.6898 ± 0.1794 | 0.4246 |
| GLCM Energy | 0.0167 ± 0.0061 | 0.0216 ± 0.0056 | < 0.0001 | 0.6758 ± 0.1932 | 0.6690 ± 0.192 | 0.4802 |
| GLCM correlation | 0.5520 ± 0.0797 | 0.5212 ± 0.0744 | < 0.0001 | 0.7554 ± 0.0983 | 0.7257 ± 0.0930 | < 0.0001 |
| GLCM ASM | 0.0003 ± 0.0004 | 0.0005 ± 0.0003 | < 0.0001 | 0.4952 ± 0.2263 | 0.4852 ± 0.2240 | 0.4650 |